\documentclass[aps,pre,reprint,groupedaddress,showpacs]{revtex4-1}

\usepackage{graphicx}
\usepackage{color} 

\begin{document}


\title{Shape evolution of numerically obtained subaqueous barchan dunes\\
\textnormal{Accepted manuscript for Physical Review E, 101, 012905 (2020), DOI: 10.1103/PhysRevE.101.012905}}



\author{Carlos A. Alvarez}
 \email{calvarez@fem.unicamp.br}
\author{Erick M. Franklin}
 \email{franklin@fem.unicamp.br}
 \thanks{Corresponding author}
\affiliation{%
School of Mechanical Engineering, UNICAMP - University of Campinas,\\
Rua Mendeleyev, 200, Campinas, SP, Brazil\\
}%


\date{\today}

\begin{abstract}

In the realm of granular bedforms, barchan dunes are strong attractors that can be found in rivers, terrestrial deserts and other planetary environments. These bedforms are characterized by a crescentic shape, which, although robust, presents different scales according to the environment they are in, their length scale varying from the decimeter under water to the kilometer on Mars. In addition to the scales of bedforms, the transport of grains presents significant differences according to the nature of the entraining fluid, so that the growth of barchans rests to be fully understood. Given the smaller length and time scales of the aquatic case, subaqueous barchans are the ideal object to study the growth of barchan dunes. In the present paper, we reproduce numerically the experiments of Alvarez and Franklin [Phys. Rev. E 96, 062906 (2017) and Phys. Rev. Lett. 121, 164503 (2018)] on the shape evolution of barchans from their initiation until they have reached a stable shape. We computed the bed evolution by using computational fluid dynamics - discrete element method (CFD-DEM), where we coupled DEM with large eddy simulation (LES) for the same initial and boundary conditions of experiments, performed in a closed-conduit channel where initially conical heaps evolved to single barchans under the action of a water flow in turbulent regime. Our simulations captured well the evolution of the initial pile toward a barchan dune in both the bedform and grain scales, with the same characteristic time and lengths observed in experiments. In addition, we obtained the local granular flux and the resultant force acting on each grain, the latter not yet previously measured nor computed. This shows that the present method is appropriate to numerical computations of bedforms, opening new possibilities for accessing data that are not available from current experiments.

\end{abstract}


\maketitle


\section{\label{sec:Intro} INTRODUCTION}

Under the action of a fluid flow, granular beds may give rise to ripples and dunes. Among the different bedforms, barchan dunes are strong attractors that can be found in highly diverse environments such as rivers, oil pipelines, open channels, terrestrial deserts, and even other planetary environments \cite{Bagnold_1, Hersen_1, Claudin_Andreotti, Parteli2}. Barchans are usually formed when grains are transported as bed load by a one-directional fluid flow, and are characterized by a crescentic shape with horns pointing downstream \cite{Bagnold_1, Hersen_1}. Although robust, this shape presents different scales according to the environment the barchans are in, varying from the decimeter and minute under water \cite{Franklin_2, Franklin_8} to the kilometer and millennium on Mars \cite{Claudin_Andreotti, Parteli2}.

Besides the bedform scales, the transport of grains is different according to the nature of the entraining fluid. In liquids, bed load is characterized by grains that move by rolling and sliding over each other \cite{Penteado}, and occasionally by small jumps with distances of the order of a few grain diameters \cite{Bagnold_3, Bagnold_4}, while in gases bed load is characterized by ballistic flights over distances much larger than the grain diameter \cite{Bagnold_1, Bagnold_3}. Because of shape and transport differences, a general explanation for the growth of barchans is still lacking.

Given the smaller length and time scales of the aquatic case, subaqueous barchans are the ideal object to study the growth of barchan dunes. For this reason, many studies carried out experiments in water tanks and channels in order to obtain the length and time scales of barchans \cite{Hersen_1, Endo, Franklin_8, Hori, Alvarez} and typical trajectories and velocities of moving grains \cite{Alvarez3, Wenzel, Alvarez4}. The smaller and faster scales of the subaqueous case have allowed the establishment of relations and scaling factors crucial to understand the dynamics of much larger and slower barchans found in other environments \cite{Hersen_1, Claudin_Andreotti, Parteli2}.

Another way to investigate the dynamics of barchans is by carrying out numerical simulations. The first numerical works computing large scale dunes employed continuum models for the grains \cite{Sauermann_4, Herrmann_Sauermann, Kroy_A, Kroy_C, Kroy_B, Schwammle, Parteli4} and considered that saltation was the mode of entrainment, occurring mainly in the longitudinal direction, with some transverse diffusion. Therefore, those models are suitable to simulations of aeolian and Martian dunes consisting of a large number of salting grains, in which, indeed, they succeeded. However, the transverse displacements of grains in the subaqueous case are, in average, comparable to the longitudinal ones, as shown recently by Refs. \cite{Alvarez3, Wenzel, Alvarez4}, a situation different from what is generally conjectured for the aeolian and Martian cases. If we consider that subaqueous barchans measure a few decimeters, consisting of only some tens of thousand grains, and that saltation is not the preferable mode of entrainment, then continuous models are not well justified in the aquatic case, although they may give reasonable results if the transverse diffusion is tuned appropriately or if bed load is considered to follow more closely the fluid. For example, Khosronejad and Sotiropoulos \cite{Khosronejad} succeeded in computing fields of subaqueous barchans by using a continuum model for the grains. They coupled a large eddy simulation (LES) model for the water and water-suspension mixture with a continuum model involving the entrainment and Exner equations for the granular bed, obtaining morphological characteristics of bedforms during the evolution of an initially flat bed toward a barchan field. However, the dynamics at the grain scale is not accessible with this method.

More recently, simulations of bed load as a discrete medium by using Euler-Lagrange methods, such as computational fluid dynamics - discrete element method (CFD-DEM), succeeded in capturing the main features of bed load \cite{Schmeeckle, Kidanemariam2, Liu, Sun, Pahtz_3} and bedforms \cite{Kidanemariam, Kidanemariam3}. Schmeeckle \cite{Schmeeckle} presented a model coupling DEM with LES based on the open source codes \mbox{CFDEM} \cite{Goniva} (www.cfdem.com), OpenFOAM (www.openfoam.org) and LIGGGHTS \cite{Kloss, Berger}, that was used to compute sand transport by water. The author carried out computations for different flow strengths over initially flat beds, being interested essentially in the entrainment of grains by the water flow. His results showed that bed-load grains move generally in contact with the bed and that saltation is of lesser importance in the subaqueous case, with most of grains being dislodged directly by the water flow. P\"ahtz and Dur\'an \cite{Pahtz_3} investigated numerically the mechanisms of the entrainment of grains by using DEM coupled with Reynolds-averaged Navier-Stokes equations (RANS). The authors examined the conditions for entrainment directly by the fluid (fluid entrainment) and particle-bed impacts (impact entrainment), and found that fluid entrainment is important only for viscous cases, the impact entrainment being more important in all the other cases. For a given range of flows, Refs. \cite{Schmeeckle} and \cite{Pahtz_3} present contrasting results with respect to grain entrainment.

Kidanemariam and Uhlmann \cite{Kidanemariam2, Kidanemariam, Kidanemariam3} numerically investigated bed load and the growth of bedforms in the subaqueous case. The authors used direct numerical simulations (DNS) for the fluid, DEM for the grains, and immersed boundary (IB) for the coupling between fluid and grains. Their simulations showed that bed load is computed properly by the proposed method, and wavelengths, amplitudes and celerities of bedforms are well captured, being in agreement with experimental data available in the literature. The proposed methodology is currently the most accurate, avoiding the use of turbulence models for the fluid, capturing all the scales of turbulence down to Kolmogorov scale, and fully solving the flow around each grain. However, the computational cost is exceedingly high, so that the time required for obtaining developed bedforms is seldom reached \cite{Colombini}.

Although some numerical investigations succeeded in capturing the formation of subaqueous barchans, they did not report detailed measurements in the bedform and grain scales, namely the evolution of lengths and times of bedforms and the trajectories of individual grains over barchans. In addition, the role of water in the direct entrainment of grains is still object of debate. Detailed morphological evolution, local characteristics of the water flow and trajectories of individual grains obtained from numerical computations are important to evaluate the suitability of numerical methods and provide information not accessible from reported experiments, such as the mode of entrainment and resultant forces on grains.

\subsection{\label{sec:Prior work} Prior work}

In previous studies \cite{Alvarez, Alvarez3}, we investigated experimentally the morphology and the trajectories of grains migrating to horns during the growth of subaqueous barchans. The experiments were conducted in a closed-conduit channel where initially conical heaps evolved to single barchans under the action of a turbulent water flow. The bedforms consisted of glass beads with diameters within 0.15 mm $\leq\,d\,\leq$ 0.25 mm and 0.40 mm $\leq\,d\,\leq$ 0.60 mm, and the cross-sectional mean velocities of water $U$ were 0.243, 0.294 and 0.364 m/s, corresponding to shear velocities on the bottom wall of the channel $u_*$ of 0.0141, 0.0168 and 0.0202 m/s, respectively. Based on the experimental results, we found the characteristic times $0.5 t_c$ and $2.5 t_c$ for the growth and equilibrium of barchans, respectively, where $t_c$ is a proposed time scale computed as the length of the bedform divided by its celerity, and that most of grains migrating to horns during the growth of barchans were initially on upstream regions at the periphery of the initial pile, the grains traveling with significant transverse displacements.

\subsection{\label{sec:Current work} This study}    

In this paper, we reproduce numerically some of the experimental results of Alvarez and Franklin \cite{Alvarez, Alvarez3} on the shape evolution of barchans, from their initiation until they have reached a stable shape, and the trajectories of grains migrating to horns. Besides the evolution of the bedform morphology and the trajectory of grains, we obtained numerically the local granular flux and the resultant force acting on each grain, for which data are still missing in the literature, and the local water flow over a single barchan, which we compare with the experiments of Charru and Franklin \cite{Franklin_11}. Our simulations were performed by using CFD-DEM to compute numerically the formation of single barchans from initially conical piles, where the fluid was computed with LES and the grains by solving linear and angular momentum equations applied to each particle. LES, although needing subgrid turbulence models, is able to compute the flow around bedforms including recirculation regions and some large turbulence structures, with a much lower computational cost than DNS. Our simulations captured well not only the formation of barchans, but also their length scale, the time scale $t_c$ for the growth of horns, the trajectories of grains migrating to horns, the local flux of grains, and the resultant force acting on each grain. Our results show that the used method is appropriate to numerical computations of subaqueous bedforms, opening new possibilities for accessing data not available from current experiments.

In the following, Secs. \ref{sec:model} and \ref{sec:setup} present, respectively, the model equations and numerical setup, and Sec. \ref{sec:Res} presents the results for the bedform morphology, local characteristics of the water flow over developed barchans, trajectory of individual grains, local flux of grains, and resultant force acting on each grain. Section \ref{sec:Conclu} presents the conclusions.

\section{\label{sec:model} MODEL DESCRIPTION}

Our numerical investigation was conducted with {CFD-DEM}, where the dynamics of each individual particle was computed by DEM \cite{Cundall} using the resulting forces and torques on each particle, the fluid flow was computed by LES, and momentum coupling was made between solids and fluid.

\subsection{\label{sec:solids} Grains}

The solid particles are treated in a Lagrangian framework, with the dynamics of each particle being computed by the linear and angular momentum equations, given by Eqs. \ref{Fp} and \ref{Tp}, respectively,

\begin{equation}
m_{p}\frac{d\vec{u}_{p}}{dt}= \vec{F}_{fp} + \vec{F}_{c} + m_{p}\vec{g}
\label{Fp}
\end{equation}

\begin{equation}
I_{p}\frac{d\vec{\omega}_{p}}{dt}=\vec{T}_{c}
\label{Tp}
\end{equation}

\noindent where $\vec{g}$ is the acceleration of gravity and, for each particle, $m_{p}$ is the mass, $\vec{u}_{p}$ is the velocity, $I_{p}$ is the moment of inertia, $\vec{\omega}_{p}$ is the angular velocity, $\vec{F}_{fp}$ is the resultant of fluid forces on grains, $\vec{F}_{c}$ is the resultant of contact forces between solids, and $\vec{T}_{c}$ is the resultant of contact torques between solids. In the balance of angular momentum, Eq. \ref{Tp}, we neglect momentum variations caused by the fluid because the term due to contacts is much higher \cite{Tsuji, Tsuji2, Liu}. In the present study, we consider that the resultant of fluid forces acting on each particle is made up of components given by the fluid drag, fluid stresses and added mass:

\begin{equation}
\vec{F}_{fp} = \vec{F}_{D} + \vec{F}_{stress} + \vec{F}_{am}
\label{Ff}
\end{equation}

\noindent where $\vec{F}_{D}$ is the drag force caused by the fluid on grains, $\vec{F}_{stress}$ = $V_{p}\left[ -\nabla P + \nabla \cdot\vec{\vec{\tau}} \right]$ is the force caused by fluid stresses and $\vec{F}_{am}$ is the added mass force, $V_{p}$ being the volume of one solid particle, $P$ the fluid pressure and $\vec{\vec{\tau}}$ the deviatoric stress tensor of the fluid. In Eq. \ref{Ff}, we neglect the Basset, Saffman and Magnus forces as they are usually considered of lesser importance in CFD-DEM simulations \cite{Zhou}.

The contact forces and torques are the result of contacts between particles and between particles and the wall. They can be expressed, respectively, by Eqs. \ref{Fc} and \ref{Tc},

\begin{equation}
\vec{F}_{c} = \sum_{i\neq j}^{N_c} \left(\vec{F}_{c,ij} \right) + \sum_{i}^{N_w} \left( \vec{F}_{c,iw} \right)
\label{Fc}
\end{equation}

\begin{equation}
\vec{T}_{c} = \sum_{i\neq j}^{N_c} \vec{T}_{c,ij} + \sum_{i}^{N_w} \vec{T}_{c,iw}
\label{Tc}
\end{equation}

\noindent where $\vec{F}_{c,ij}$ and $\vec{F}_{c,iw}$ are the contact forces between particles and between particles and the wall, respectively, $\vec{T}_{c,ij}$ is the torque due to the tangential component of the contact force between particles $i$ and $j$, and $\vec{T}_{c,iw}$ is the torque due to the tangential component of the contact force between particle $i$ and the wall, $N_c$ - 1 being the number of particles in contact with particle $i$, and $N_w$ the number of particles in contact with the wall.

\subsection{\label{sec:fluid} Fluid}

The fluid is the continuous phase and is treated in an Eulerian framework. For an incompressible flow, mass and momentum equations are given by Eqs. \ref{mass} and \ref{mom}, respectively,

\begin{equation}
\nabla\cdot\vec{u}_{f}=0
\label{mass}
\end{equation}

\begin{equation}
\frac{\partial{\rho_{f}\vec{u}_{f}}}{\partial{t}} + \nabla \cdot (\rho_{f}\vec{u}_{f}\vec{u}_{f}) = -\nabla P + \nabla\cdot \vec{\vec{\tau}} + \rho_{f}\vec{g} - \vec{f}_{fp}
\label{mom}
\end{equation}

\noindent where $\vec{u}_{f}$ is the fluid velocity, $\rho_{f}$ is the fluid density (equal to 10$^{3}$ kg/m$^3$ in the present case), and $\vec{f}_{fp}$ is the resultant of fluid forces acting on each grain, $\vec{F}_{fp}$, by unit of fluid volume, representing the momentum transfer from the fluid to the solids.

\section{\label{sec:setup} NUMERICAL SETUP}

In the present study, we used the open source code \mbox{CFDEM} \cite{Goniva} (www.cfdem.com) for our CFD-DEM computations. \mbox{CFDEM} links the open source code OpenFOAM, which computes the fluid flow based on the finite volume method (FVM), with the open source code LIGGGHTS \cite{Kloss, Berger}, which computes the dynamics of grains.

Concerning the grains, we considered a Hertzian model for which we set up the coefficient of restitution $e$ as approximately 0.1 because collisions are expected to be viscous damped \cite{Schmeeckle}. The friction coefficient $\mu_{fr}$ was considered as 0.6 \cite{Schmeeckle, Liu}, and the Young's modulus $E$ and Poisson ratio $\sigma$ were obtained from Refs. \cite{Tsuji, Tsuji2}. The Young's modulus used in the simulations are considerable smaller than real values in order to reduce the DEM time step without affecting significantly the numerical outputs \cite{Tsuji2}. The main parameters used in our simulations are listed in Tab. \ref{tabsim}.

\begin{table}[ht]
	\begin{center}
	\caption{DEM parameters}
	\begin{tabular}{c c}
		\hline\hline
		Particle diameter $d$ (mm)  & 0.5 \\
		Particle density $\rho_p$ (kg/m$^3$) & 2500 \\
		Young's Modulus $E$ (MPa) & 5 \\
		Poisson ratio $\sigma$ & 0.45 \\
		Restitution coefficient $e$ & 0.1 \\
		Friction coefficient $\mu_{fr}$ & 0.6\\
		Initial number of particles $N$ & 4 $\times$ 10$^4$\\
		Time step (s) & 5 $\times$ $10^{-6}$\\
		\hline
	\end{tabular}
		\label{tabsim}
    \end{center}
\end{table}

For the fluid, we used LES with subgrid stresses given by the wall-adapting local eddy-viscosity (WALE) model \cite{Nicoud}. The domain was set to 0.3 $\times$ 0.05 $\times$ 0.16 m divided into 150, 150 and 160 segments in the streamwise, $x$, wall-normal, $y$, and spanwise, $z$, directions, respectively. The dimensions in $y$ = $2\delta$ and $z$ directions are the same as the ones of previous experiments \cite{Alvarez, Alvarez3, Alvarez4}. The segments in the $x$ and $z$ directions are uniform in size, whereas in the $y$ direction they are unevenly spaced. Details of the grid and Reynolds numbers employed in our simulations are summarized in Tab. \ref{TabLES}, where $Re$ = $U 2\delta \nu^{-1}$ is the channel Reynolds number based on the cross-sectional mean velocity and $Re_*$ = $u_* \delta \nu^{-1}$ is the Reynolds number based on the shear velocity, $\nu$ being the kinematic viscosity (equal to 10$^{-6}$ m$^2$/s in the present case).

\begin{table}[ht]
	\begin{center}
		\caption{Computational grid and Reynolds numbers. $\Delta x^{+}$ and $\Delta z^{+}$ are the grid spacings $\Delta x$ and $\Delta z$ in the streamwise and spanwise directions, respectively, scaled in inner wall units, $\nu u_*^{-1}$. $\Delta y^{+}_{1}$ and $\Delta y^{+}_{c}$ are the grid spacing in the wall-normal direction at the first and center points, respectively, scaled in inner wall units. $Re$ = $U 2\delta \nu^{-1}$ and $Re_*$ = $u_* \delta \nu^{-1}$, where $U$ is the cross-sectional mean velocity.}
		\begin{tabular}{c c c}
			\hline
			 & Case 1 & Case 2 \\ \hline \hline
			$\Delta x^+ $ & 33.6 & 40.4  \\
			$\Delta y^{+}_{1} $ & 0.91 & 1.10  \\
			$\Delta y^{+}_{c} $ & 17.18 & 20.65  \\
			$\Delta z^+ $ & 16.80 & 20.20  \\
			$Re$ & 1.47$\times$10$^4$ & 1.82$\times$10$^4$\\
			$Re_*$ & 420 & 506
			\\ 
			\hline
		\end{tabular}
		\label{TabLES}
	\end{center}
\end{table}

In Eq. \ref{mom}, the term of momentum transfer between the fluid and solids is computed as $\vec{f}_{fp}$ = $\vec{F}_{fp} / V_{cell}$, where $V_{cell}$ is the volume of the considered computational cell. The drag force in Eq. \ref{Ff} is computed as in Eq. \ref{eq:FD},

\begin{equation}
\vec{F_D} =  A_p C_D  \rho_f (\vec{u_f} - \vec{u_p}) \left| \vec{u_f} - \vec{u_p}\right|
\label{eq:FD}  
\end{equation}      

\noindent where $A_p$ is the particle cross sectional area and $C_D$ is a drag coefficient obtained from the Gidaspow model \cite{Gidaspow}.

\begin{figure}[h!]
\begin{center}
		\includegraphics[width=0.9\columnwidth]{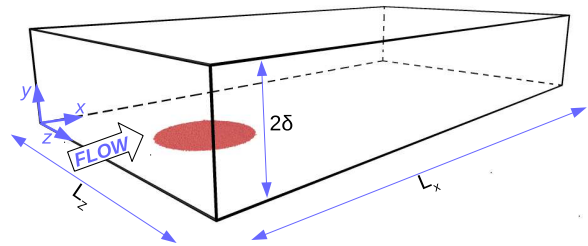}\\	
\end{center}
    \caption{Layout of the computational setup. In the simulations, $L_x$ = 0.3 m, $2\delta$ = 0.05 m and $L_z$ = 0.16 m.}
    \label{fig:num_layout}
\end{figure}

Prior to CFD-DEM simulations, single phase flows were computed in the computational domain, corresponding to a rectangular cross-sectional channel. Periodic conditions were considered in streamwise and spanwise directions, and the final realization was saved to be used as the fluid initial condition for the CFD-DEM simulations with grains.

Also prior to starting the simulations, the water flow was set to zero velocity and the grains were poured from above, falling freely in water at a longitudinal location 0.05 m from the domain inlet and centered in the transverse direction. After a period for the grains to settle, we obtained an initially conical pile with radius $R$ = 0.025 m, similar to Refs. \cite{Alvarez, Alvarez3, Alvarez4}, and center at 0.05 m from the domain inlet. Once the initial condition for the grains reached, we imposed the final realization of the single-phase LES computations as the initial condition for water. The boundary conditions for the fluid were impermeability and no-slip conditions at the top and bottom walls of the channel, and periodic conditions in the longitudinal and transverse directions. The boundary conditions for the solid particles were solid walls at the top and bottom walls, free exit at the outlet, and no mass entering at the inlet, so that the barchan decreased slightly in size while migrating, in the same manner as in Refs. \cite{Alvarez, Alvarez3, Alvarez4}. Figure \ref{fig:num_layout} shows a layout of the computational domain.

The resulting conditions for sediment transport are $s$ = 2.5, $Ga$ = 43, $Im$ = 74, and $\theta$ = 0.04 and 0.06, where $s = \rho_p/ \rho_f$ is the density ratio, $Ga = \sqrt{ \left( s-1 \right) g d^3} / \nu$ the Galileo number, $\theta = u_*^2/ \left( \left( s -1 \right) g d \right)$ the Shields number, and $Im = Ga\sqrt{s+0.5}$ the impact number proposed by P\"ahtz and Dur\'an \cite{Pahtz_3} to account for the importance of impact entrainment on bed load and saltation. In the present case 3 $\leq$ $\theta \times Im$ $\leq$ 4 and $Im$ $>$ 20, placing the bed-load conditions in the limit between fluid and impact entrainment, according to Ref. \cite{Pahtz_3}. The experiments reported in Refs. \cite{Alvarez, Alvarez3, Alvarez4} showed grains rolling and sliding following closely the mean water flow, pointing in the direction of direct entrainment by the fluid.

The tracking of bedforms and individual grains, as well as computations of morphology characteristics and grain trajectories, were carried out in the same manner as in our previous experiments \cite{Alvarez, Alvarez3, Alvarez4}, making use of scripts written for this purpose based on Refs. \cite{Kelley, Alvarez2}.

\section{\label{sec:Res} RESULTS AND DISCUSSION}

\subsection{\label{sec:res_single} Single water flow}

Computations of single-phase channel flows were necessary to be used as initial condition for the simulations with solid particles. In order to evaluate our LES computations, the single phase simulations for $Re_*$ = 420 were compared with the DNS results of Moser et al. \cite{Moser} for $Re_*$ = 395. For that, we averaged 1,000 velocity fields in time (the last of 3,000 fields) and in the $x$ and $z$ directions, obtaining the profiles of mean velocity and Reynolds stress, which were afterward compared with Ref. \cite{Moser}.

\begin{figure}[h!]	
	\begin{minipage}[c]{\columnwidth}
		\begin{center}
		\includegraphics[width=.85\linewidth]{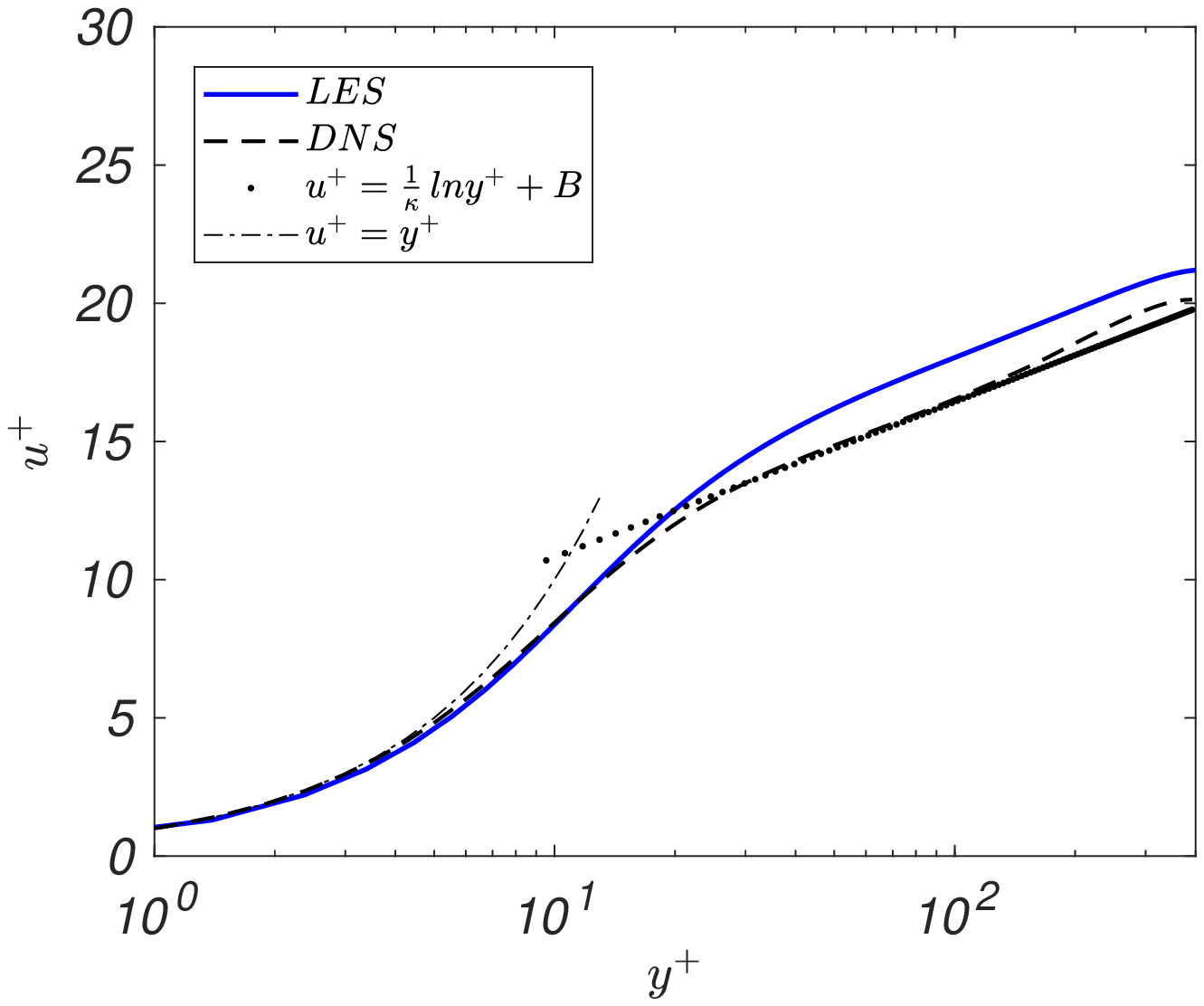}\\
			(a)
		\end{center}
	\end{minipage} \hfill
		\begin{minipage}[c]{\columnwidth}
    	\begin{center}
		\includegraphics[width=.85\linewidth]{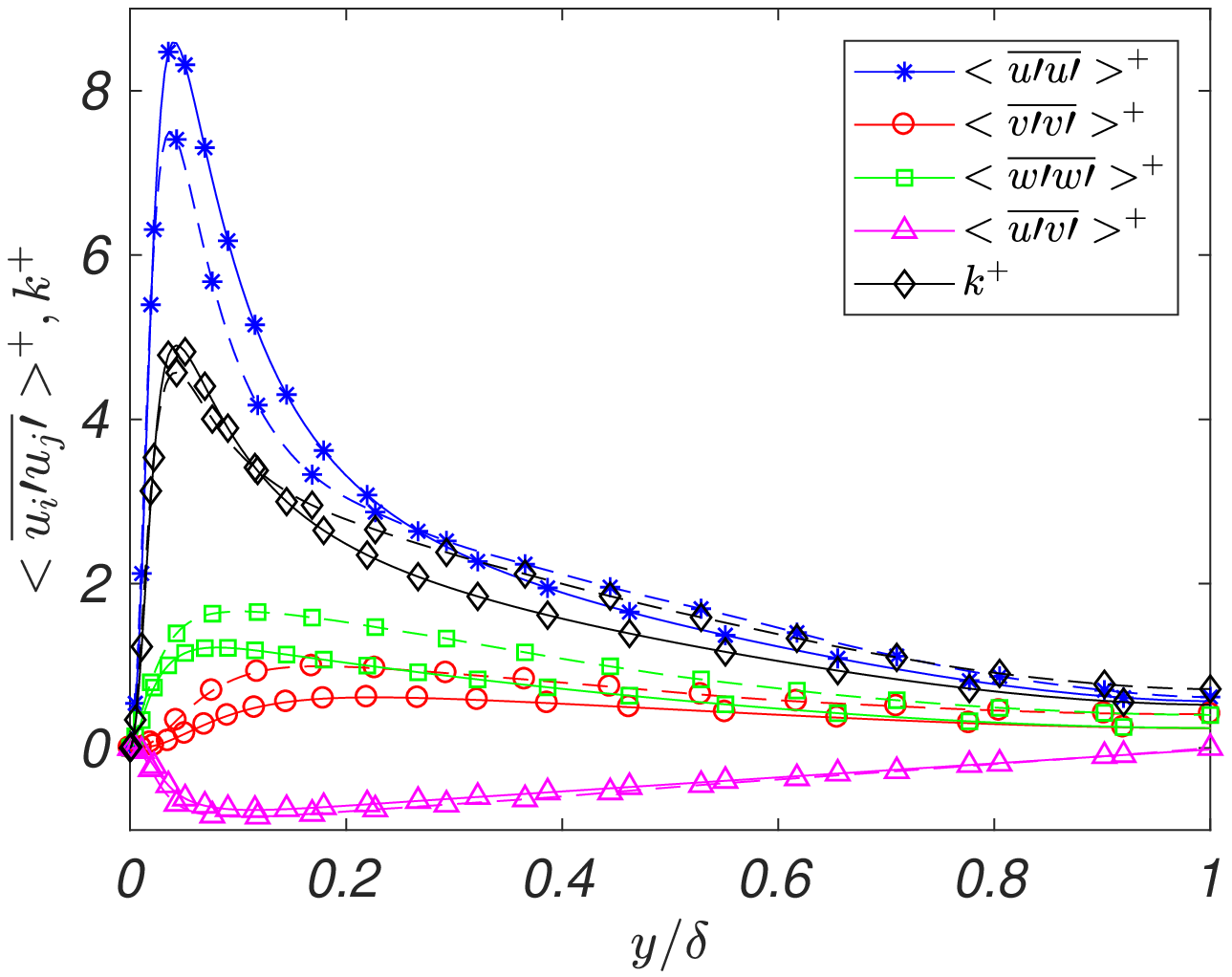}\\	
		(b)
		\end{center}
	\end{minipage}
	\caption{Single-phase flow in the channel. a) Profile of the longitudinal component of the mean velocity $u$ in the traditional log-normal scales, normalized by the inner scales. b) Profiles of the components of the Reynolds stress and turbulent kinetic energy normalized by the inner scales. Solid lines correspond to the present LES results (for $Re_*$ = 420) and dashed lines to the DNS results of Moser et al. \cite{Moser} (for $Re_*$ = 395). Figure (a) presents also the law of the wall with $\kappa$ = 0.41 and $B$  = 5.2.}
	\label{fig:single_phase}
\end{figure}

Figures \ref{fig:single_phase}(a) and \ref{fig:single_phase}(b) present, respectively, the profiles of the longitudinal component of the mean velocity in the traditional log-normal scales and of the components of the Reynolds stress, normalized by the inner scales, where solid lines correspond to the present LES results and dashed lines to the DNS results of Moser et al. \cite{Moser}. In these figures, $u$ is the longitudinal component of the mean velocity, $u^+$ = $u/u_*$, $k$ is the turbulent kinetic energy, and $u'$, $v'$ and $w'$ are the longitudinal, vertical and transverse components of velocity fluctuations, respectively. In the case of velocity fluctuations and turbulent kinetic energy, the superscript $+$ means division by $u_*^2$, overbar means temporal averages, and $<>$ refers to spatial averages in the $x$ and $z$ directions. Figure \ref{fig:single_phase}(a) presents also the law of the wall with $\kappa$ = 0.41 and $B$  = 5.2.

In Fig. \ref{fig:single_phase}(a) we can observe that in the viscous region the mean profile obtained with LES shows a good agreement with both the law of the wall and DNS results of Ref. \cite{Moser}, while in the overlap region LES results are shifted upwards with the same slope when compared with both the law of the wall and DNS. LES results could have a better agreement with DNS if we further refined the mesh; however, we are not carrying out particle-resolved simulations, so that the present results are adequate for the purpose of initial conditions for simulations with bed load. Figure \ref{fig:single_phase}(b) shows a good agreement between second order moments obtained by LES and DNS. Finally, we observe that two-dimensional measurements with particle image velocimetry (PIV) in a channel with the same dimensions provided similar results \cite{Franklin_9, Franklin_10}.

\subsection{\label{sec:res_twophase} Simulations with bed load}

\subsubsection{\label{sec:res_morphology} Morphology of bedforms}

With the grains settled in the channel, forming an initially conical pile, the output of the LES simulation for the single-phase flow was imposed as the initial condition for the fluid. Once the simulations started, with periodic conditions for the fluid in the $x$ and $z$ directions, the turbulent flow was sustained and the grains entrained by the water flow, deforming the initial pile in a barchan dune, as can be seen in Fig. \ref{fig:compar} and in a movie from one numerical simulation available as Supplemental Material (see  Supplemental Material \cite{Supplemental} for a movie showing the development of a barchan dune).

\begin{figure} [h!]
	\centering
	\includegraphics[width=0.99\columnwidth]{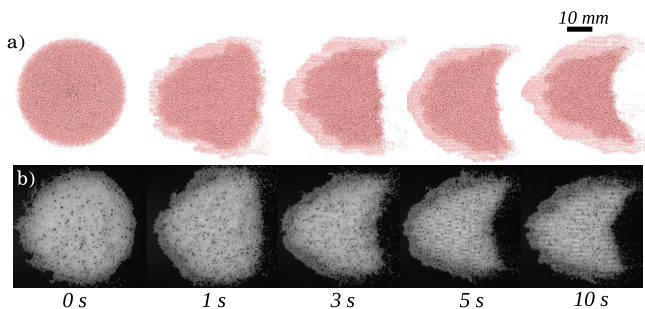}
	\caption{Top views of an initially conical heap deformed by a turbulent water flow at different times. The flow is from left to right. a) Numerical results. b) Experiments of Ref. \cite{Alvarez3}. $Re$ = 1.82$\times$10$^4$, $d$ = 0.5 mm, and $\rho_p$ = 2500 kg/m$^3$.}
	\label{fig:compar} 
\end{figure}

Figure \ref{fig:compar} shows top views of the bedform as it is deformed from a conical heap in a barchan dune by the water flow. Figure \ref{fig:compar}(b) corresponds to images from experiments for $Re$ = 1.82$\times$10$^4$, $d$ = 0.5 mm, and $\rho_p$ = 2500 kg/m$^3$, and Fig. \ref{fig:compar}(a) consists of snapshots of our numerical simulations for the same conditions. From the snapshots and Refs. \cite{Alvarez3, Supplemental}, we observe a similar evolution of bedforms in experiments and numerical simulations: once the water starts to flow, the pile is deformed, with the formation of a slip face and growing of horns. After some time, the bedform adopts a crescentic shape. The evolution of length and time scales are discussed next.

\begin{figure}[h!]
\begin{center}
	\begin{tabular}{c}
	\includegraphics[width=0.85\columnwidth]{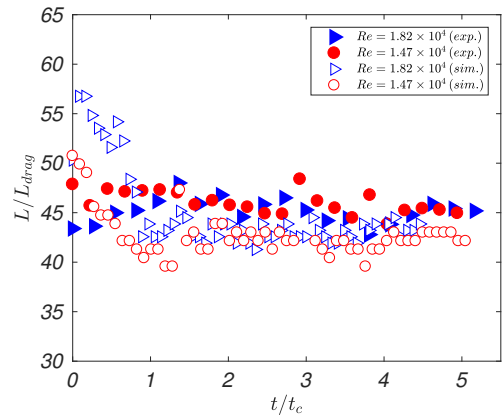}\\
	(a)\\
	\includegraphics[width=0.85\columnwidth]{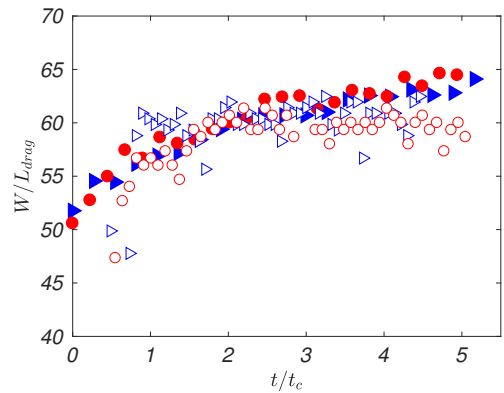}\\
	(b)
	\end{tabular} 
\end{center}
    \caption{(a) $L/L_{drag}$ versus $t/t_c$ and (b) $W/L_{drag}$ versus $t/t_c$. The open symbols correspond to numerical results, solid symbols to the experimental results of Refs. \cite{Alvarez, Alvarez3}, and Reynolds numbers are listed in the key.}
    \label{fig:L_W_time}
\end{figure}

A direct comparison between experiments and numerical simulations for the evolutions of length and width of barchans is presented in Fig. \ref{fig:L_W_time}. Figures \ref{fig:L_W_time}(a) and \ref{fig:L_W_time}(b) show, respectively, the length $L$ and width $W$ of barchans normalized by the length scale $L_{drag}$ as functions of time $t$ normalized by the timescale $t_c$. The open symbols correspond to numerical results, solid symbols to the experimental results of Refs. \cite{Alvarez, Alvarez3}, and Reynolds numbers are listed in the key of Fig. \ref{fig:L_W_time}(a). $L_{drag}$ = $(\rho_s / \rho_f )d$ is an inertial length for the aeolian case \cite{Hersen_1, Sauermann_4}, which is proportional to the length for the stabilization of sand flux, and $t_c$ = $L_{eq} / C$ is the timescale for the barchan evolution \cite{Alvarez}, where $L_{eq}$ is the barchan length once the crescentic shape is attained and $C$ is the dune celerity.

For $t$ $<$ $t_c$, the numerically obtained values are different from the experimental ones, with, in addition, a different behavior for $L$, which decreases in the numerical results while it remains constant in the experiments. The difference in the initial behavior is due to slight differences in the initial conditions of numerical simulations when compared to experiments. In their turn, part of differences in the initial conditions results from the pouring of grains on the bottom wall, where the friction coefficient between grains and wall is fixed in the numerical simulations, but is susceptible to variations and uncertainties in the experiments. In general, the bedforms in numerical simulations and experiments evolve in a similar manner, reaching roughly the same length and width for $t$ $\geq$ $t_c$.

\begin{figure}[h!]
\begin{center}
		\includegraphics[width=0.85\columnwidth]{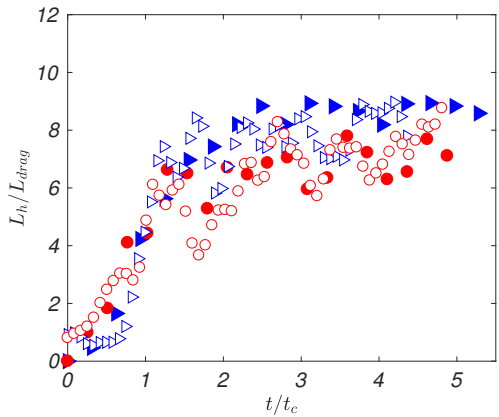}\\	
\end{center}
    \caption{$L_h/L_{drag}$ versus $t/t_c$. The open symbols correspond to numerical results, solid symbols to the experimental results of Refs. \cite{Alvarez, Alvarez3}, and Reynolds numbers are listed in the key of Fig. \ref{fig:L_W_time}(a).}
    \label{fig:horns_length}
\end{figure}

The time evolution of the length of horns $L_h$ is shown in Fig. \ref{fig:horns_length}, which presents $L_h/L_{drag}$ as a function of $t/t_c$. The open symbols correspond to numerical results, solid symbols to the experimental results of Refs. \cite{Alvarez, Alvarez3}, and Reynolds numbers are listed in the key of Fig. \ref{fig:L_W_time}(a). In the case of $L_h$, the numerical results show a good agreement with experiments, the length increasing with the same slope for $t$ $<$ $2.5 t_c$ and reaching the same constant value for $t$ $\geq$ $2.5 t_c$.

Arguing that the main feature of a barchan dune is its horns, Alvarez and Franklin \cite{Alvarez} proposed, based on the evolution of horns, that an initially conical pile deforms under the action of a water flow and reaches a stable crescentic shape when $t$ = $2.5 t_c$. For $t$ $\geq$ $2.5 t_c$, the subaqueous barchan migrates keeping the same crescentic shape, being referred in Ref. \cite{Alvarez} as stable dunes. Because our numerical simulations predict the same evolution for horns at all times, and similar behaviors for the dune length and width for $t$ $\geq$ $t_c$, we consider that the present simulations capture well the morphology of evolving barchans.

\subsubsection{\label{sec:res_flow} Flow over single barchans}

\begin{figure}[h!]
\begin{center}
		\includegraphics[width=0.99\columnwidth]{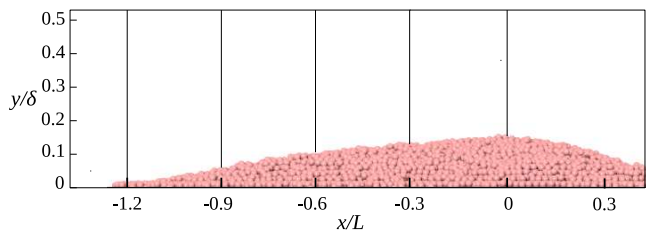}\\	
\end{center}
    \caption{Side view of a barchan dune, showing the adopted coordinate system for the flow over the dune. Vertical lines correspond to the longitudinal positions of the profiles plotted in Figs. \ref{fig:uu_uv} and \ref{fig:uv}.}
    \label{fig:side_view_dune}
\end{figure}

Second order moments for the flow over a barchan dune were obtained by time averaging 10,000 instantaneous fields (the last of 20,000 fields) in the vertical symmetry plane of the dune. Because the flow evolve along the dune, we present next only some profiles along the stoss side until reaching the dune crest. In order to obtain a significant number of fields while allowing a direct comparison with the PIV experiments reported in Charru and Franklin \cite{Franklin_11}, we artificially fixed the barchan dune by increasing considerably the density of grains at some point after $t$ = $2.5 t_c$. This procedure is similar to the one of Ref. \cite{Franklin_11}, where the fluid flow was reduced after $t$ = $2.5 t_c$ in order to remain slightly below the threshold for bed load. Neglect bed load in the subaqueous case is justified by its smaller effect on the water (feedback effect) when compared with the shape perturbation \cite{Franklin_11, Franklin_9}. We adopted the same coordinate system as in Ref. \cite{Franklin_11}, with the origin of the longitudinal coordinate at the dune crest, as shown in Fig. \ref{fig:side_view_dune}, and the use of the displaced coordinate $y_d$ = $y - h(x)$, where $h(x)$ is the local height of the barchan in its vertical symmetry plane.

\begin{figure}[h!]
\begin{center}
	\begin{tabular}{c}
	\includegraphics[width=0.85\columnwidth]{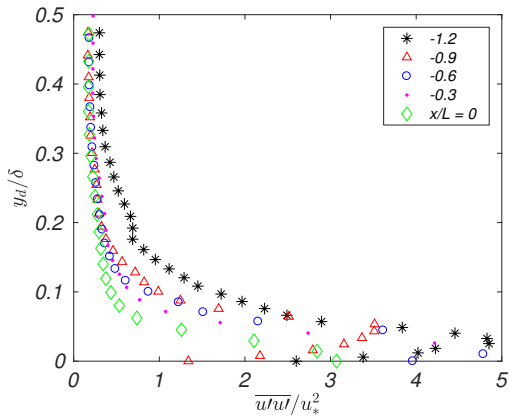}\\
	(a)\\
	\includegraphics[width=0.85\columnwidth]{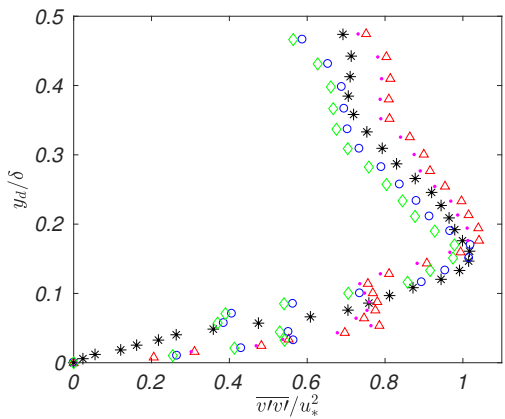}\\
	(b)
	\end{tabular} 
\end{center}
    \caption{Profiles of (a) $\overline{u'u'}$ and (b) $\overline{v'v'}$ normalized by $u_*^2$, using the displaced coordinate $y_d$ normalized by $\delta$. Each different symbol corresponds to a longitudinal position normalized by $L$, with origin at the dune crest, and listed in the key.}
    \label{fig:uu_uv}
\end{figure}

\begin{figure}[h!]
\begin{center}
		\includegraphics[width=0.85\columnwidth]{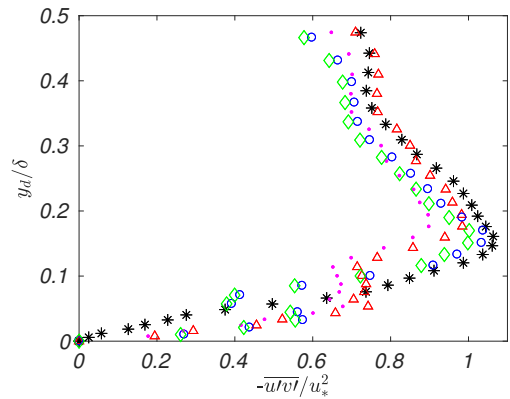}\\	
\end{center}
    \caption{Profiles of $-\overline{u'v'}$ normalized by $u_*^2$, using the displaced coordinate $y_d$ normalized by $\delta$. Each different symbol corresponds to a longitudinal position normalized by $L$, with origin at the dune crest, and listed in the key of Fig. \ref{fig:uu_uv}(a).}
    \label{fig:uv}
\end{figure}

Figures \ref{fig:uu_uv} and \ref{fig:uv} show the second order moments for the flow over the barchan dune. Figures \ref{fig:uu_uv}(a), \ref{fig:uu_uv}(b) and \ref{fig:uv} present some profiles of $\overline{u'u'}$, $\overline{v'v'}$ and $-\overline{u'v'}$, respectively, for different longitudinal positions. In these figures, the abscissa corresponds to $\overline{u'u'}$, $\overline{v'v'}$ or $-\overline{u'v'}$ normalized by $u_*^2$, the ordinate to the displaced coordinate $y_d$ normalized by $\delta$, and each different symbol to the longitudinal position normalized by $L$ with origin at the dune crest. The plotted positions are listed in the key of Fig. \ref{fig:uu_uv}(a).

The profiles of Figs. \ref{fig:uu_uv} and \ref{fig:uv} are very much alike to Figs. 16 and 17(b) of Ref. \cite{Franklin_11}. In particular, we note that the maximum value of $-\overline{u'v'}$ is attained within -1.2 $<$ $x/L$ $<$ -0.6, reaching values of approximately 1.1$u_*^2$, just as measured experimentally. In summary, the evolution of second order moments along the dune obtained from our LES simulations are in good agreement with the results of Ref. \cite{Franklin_11}, showing that the numerical setup adopted captures the mean features associated with the turbulent flow over the dune.

\subsubsection{\label{sec:res_bedload} Trajectories of grains}

In Alvarez and Franklin \cite{Alvarez3} we presented the trajectories of grains migrating to the horns of barchans during their growth from conical piles that consisted of 0.40 mm $\leq\,d\,\leq$ 0.60 mm glass beads. We present next the trajectories of grains that migrated to the growing horns of barchans obtained from numerical simulations. In the simulations, we tracked only grains moving over the dune (over a granular bed), because grains moving directly over the channel wall presented straighter trajectories when compared with experiments. This difference may be connected to the microstructure of the grain and wall surfaces, which are smoother in the numerical simulations, and, therefore, affect strongly grains rolling and sliding directly over the bottom wall. In the experiments of Ref. \cite{Alvarez3}, we expect that the channel wall was subject to small scratches caused by the grains in addition to other sources of microstructure uncertainties.

\begin{figure}[h!]
\begin{center}
		\includegraphics[width=0.6\columnwidth]{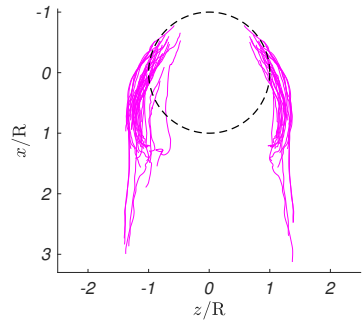}\\	
\end{center}
    \caption{Trajectories of some of the grains that migrated to horns during the growth of a barchan dune. The water flow is from top to bottom. $Re$ = $1.82 \times 10^4 $ and the dashed black circle represents the initial pile of radius R.}
    \label{fig:trajectories}
\end{figure}

Figure \ref{fig:trajectories} shows the pathlines of some grains that migrated to horns during the growth of a barchan dune from an initially conical pile for the numerical simulations with $Re$ = $1.82 \times 10^4 $. The abscissa and ordinate correspond, respectively, to the transverse and streamwise coordinates, $z$ and $x$, normalized by $R$. In Fig. \ref{fig:trajectories}, the dashed black circle displays the initial pile. We observe a behavior similar to that noted in Ref. \cite{Alvarez3}, with grains that migrate to horns describing circular paths, which implies significant transverse components. As verified experimentally by Ref. \cite{Alvarez3}, these grains come from regions upstream of the dune centroid, moving around the central region of the barchan before reaching the horns.

\begin{figure}[h!]
\includegraphics[width=0.85\linewidth]{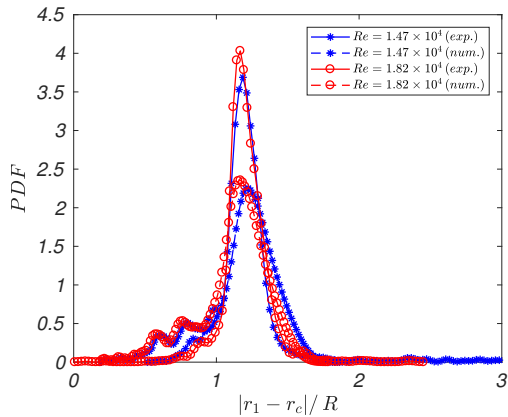}
\caption{PDFs of the original positions of the grains that migrated to horns during the growth of a barchan dune, obtained from numerical simulations and the experiments of Ref. \cite{Alvarez3}. A kernel smoothing function was used in these plots \cite{Bowman_Azzalini}.}
	\label{fig:radial_origin_horns}
\end{figure}

\begin{figure}[h!]
\begin{center}
		\includegraphics[width=0.95\columnwidth]{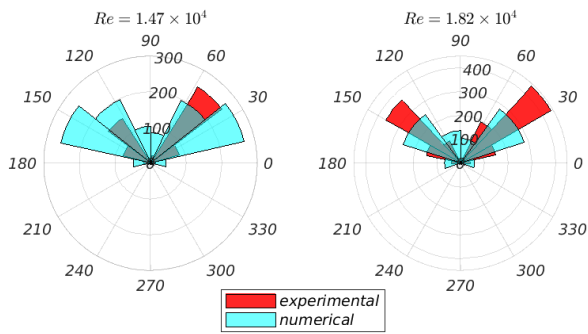}\\	
\end{center}
    \caption{Frequencies of occurrence of the initial positions of grains migrating to horns as functions of the angle with respect to the transverse direction (water flow direction is 270$^{\circ}$), obtained from numerical simulations and the experiments of Ref. \cite{Alvarez3}. The tips of appearing horns point to angles of approximately 240$^{\circ}$ and 300$^{\circ}$.}
    \label{fig:angular_origin_horns}
\end{figure}

Alvarez and Franklin \cite{Alvarez3} used radial and angular coordinates with origin at the dune centroid in order to find the original positions of grains migrating to the growing horns, and afterward computed their probability density functions (PDFs) and frequencies of occurrence. We make use here of the same method, in which we localize in polar coordinates the grains migrating to horns. The polar coordinates were defined as $(|r_1 - r_c|,\,\phi )$, in the same way as in Ref. \cite{Alvarez3}, $r_1$ being the original position of grains, $r_c$ the instantaneous position of the dune centroid, and $\phi$ the angle with respect to the transverse direction.

Figures \ref{fig:radial_origin_horns} and \ref{fig:angular_origin_horns} present the PDFs of the initial positions of grains migrating to horns as functions of $|r_1 - r_c|/R$ and their frequencies of occurrence as functions of $\phi$, respectively, for both our numerical results and the experiments of Ref. \cite{Alvarez3}. The PDFs of Fig. \ref{fig:radial_origin_horns} were computed using a kernel smoothing function \cite{Bowman_Azzalini} and the water flow direction in Fig. \ref{fig:angular_origin_horns} is 270$^{\circ}$. Figures \ref{fig:radial_origin_horns} and \ref{fig:angular_origin_horns} show a good agreement between experiments and numerical simulations, most of grains that migrate to horns being originally within $|r_1 - r_c|/R$ $>$ 1 and 15$^{\circ}$ $\leq\,\phi\,\leq$ 70$^{\circ}$ and 120$^{\circ}$ $\leq\,\phi\,\leq$ 170$^{\circ}$, i.e., at upstream regions on the periphery of the pile.

\begin{figure}[h!]
	\begin{minipage}[c]{\columnwidth}
		\begin{center}
			\includegraphics[width=.85\linewidth]{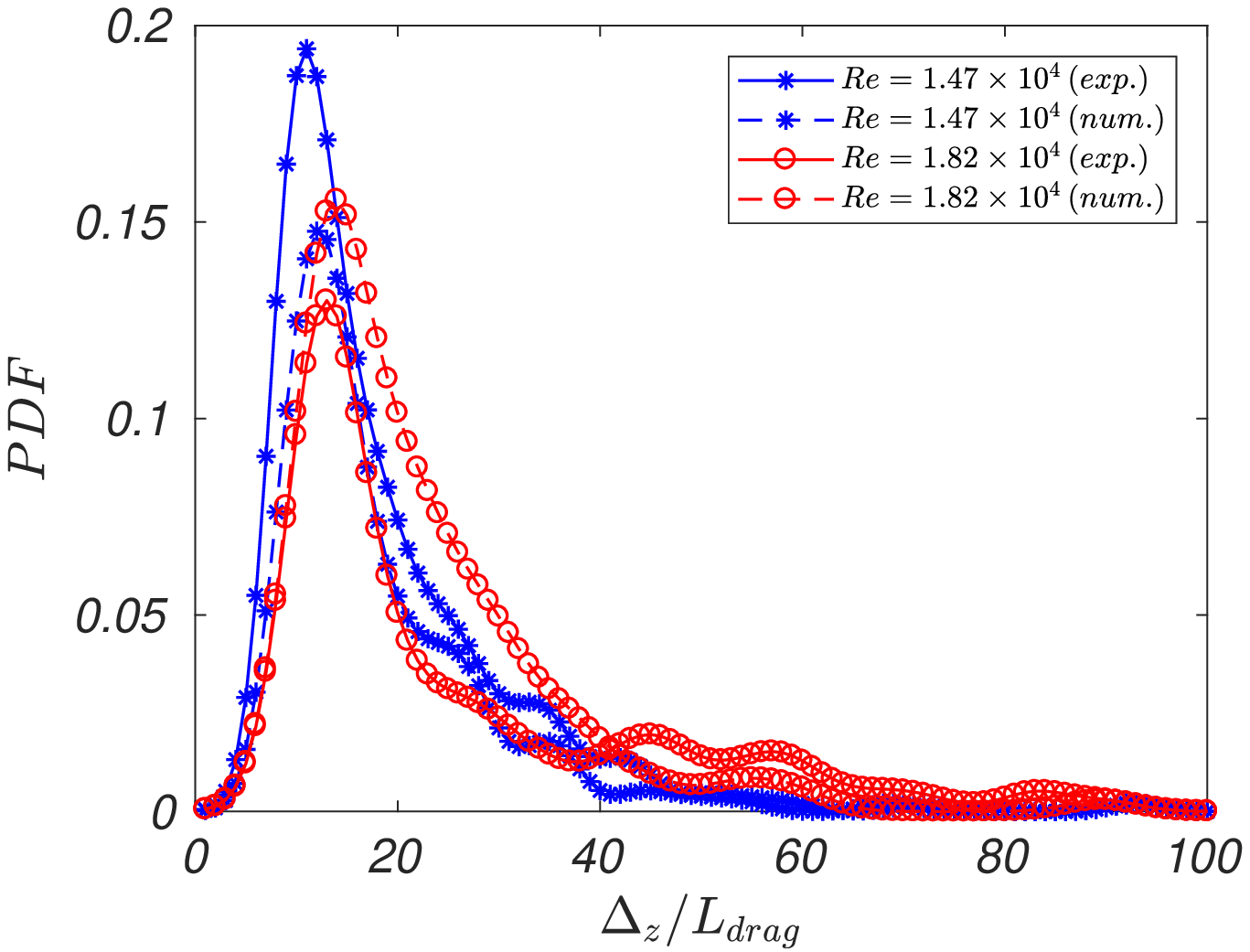}\\
			(a)
		\end{center}
	\end{minipage} \hfill
	\begin{minipage}[c]{\columnwidth}
		\begin{center}
			\includegraphics[width=.85\linewidth]{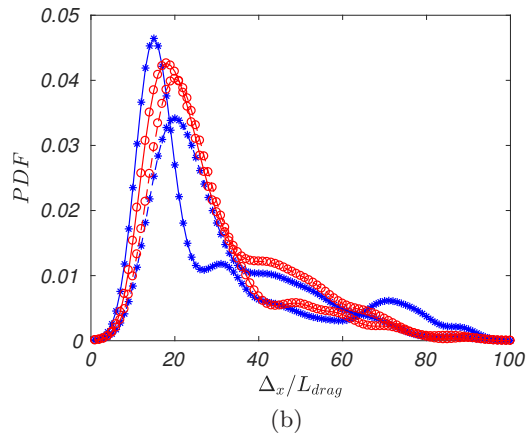}\\	
			(b)
		\end{center}
	\end{minipage}
\caption{PDFs of the total (a) transverse and (b) longitudinal distances, $\Delta_z$ and $\Delta_x$, normalized by $L_{drag}$ for numerical and experimental results, as listed in the key of figure (a).}
	\label{fig:dist_deltas}
\end{figure}

We computed the total transverse ($\Delta_z$) and longitudinal ($\Delta_x$) distances traveled by the grains that migrated to horns. Figures \ref{fig:dist_deltas}(a) and \ref{fig:dist_deltas}(b) present PDFs of $\Delta_z /L_{drag}$ and $\Delta_x /L_{drag}$, respectively, obtained from both the present simulations and the experiments of Ref. \cite{Alvarez3}, and the cases are listed in the key of Fig. \ref{fig:dist_deltas}(a). For both the numerical simulations and experiments, we observe a significant transverse component in the movement of grains, corroborating the conclusion of Ref. \cite{Alvarez3} that in the subaqueous case grains migrating to horns describe circular paths with important transverse components. By considering the most probable values of the distributions, we find 11 $\leq\,\Delta_z /L_{drag}\,\leq$ 13 and $\Delta_z / \Delta_x$ $\approx$ 0.5 for the simulations, close to the values 10 $\leq\,\Delta_z /L_{drag}\,\leq$ 12 and $\Delta_z / \Delta_x$ $\approx$ 0.5 found for the experiments.

\subsubsection{\label{sec:mass_flux} Granular flux}

We present next the flux of grains along the subaqueous barchan, an information still needing investigation even experimentally. For that, we computed the total number of grains crossing certain barchan cross sections ($x$ planes) during specific periods. By using relatively short periods, we obtained \textit{quasi-instantaneous} values for the flux.

\begin{figure}[h!]
\begin{center}
		\includegraphics[width=0.85\columnwidth]{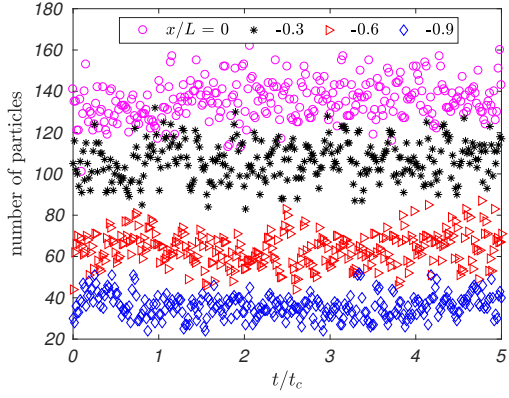}\\	
\end{center}
    \caption{Number of grains crossing certain barchan cross sections as a function of the normalized time $t/t_c$. The longitudinal positions of each considered cross section ($x$ planes) are listed in the figure key (and shown in Fig. \ref{fig:side_view_dune}), and $Re$ = 1.47$\times$10$^4$. For these computations, a time interval of 0.13 s was used.}
    \label{fig:granular_flux}
\end{figure}

Figure \ref{fig:granular_flux} presents the number of grains crossing certain barchan cross sections as a function of the normalized time $t/t_c$, where an interval of 0.13 s was used in the computations, corresponding to 0.016$t_c$. Values presented in the ordinate can be converted to g by multiplying them by 1.64 $\times$ 10$^{-4}$, which corresponds to the grain mass in g. In spite of considerable dispersion, Fig. \ref{fig:granular_flux} shows that, locally, there seems to exist a transient in the flux of grains for $t/t_c$ $\lesssim$ 1.5 and afterward the flux is roughly constant. This is, in some measure, consistent with the characteristic time $t/t_c$ = 2.5 for the growth of barchans proposed by Ref. \cite{Alvarez}. In space, Fig. \ref{fig:granular_flux} shows that the flux is not saturated along the barchan, increasing toward the crest. Although it is known that the largest value should occur just upstream the dune crest \cite{Engelund_Fredsoe}, we are not able to show that for the moment given the relatively high dispersion in obtained fluxes.

\subsubsection{\label{sec:forces_on_grains} Forces on grains}

The present simulations provide us with the resultant force acting on each grain, $\vec{F}_{p}$, corresponding to the RHS of Eq. \ref{Fp}, i.e.,  $\vec{F}_{p}$ = $\vec{F}_{fp} + \vec{F}_{c} + m_{p}\vec{g}$. This information is difficult to access experimentally, and it is not available from current experiments. In this section we analyze the resultant force on grains along the barchan dune.

\begin{figure}[h!]
\begin{center}
		\includegraphics[width=0.85\columnwidth]{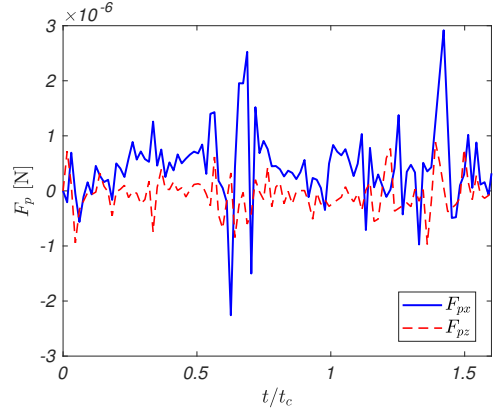}\\	
\end{center}
    \caption{Longitudinal and transverse components of the resultant force on a grain, $F_{px}$ and $F_{pz}$, respectively, as a function of time. This specific grain was entrained from an upstream region toward the crest, and $Re$ = 1.47$\times$10$^4$.}
    \label{fig:tracking_force}
\end{figure}

Figure \ref{fig:tracking_force} presents the resultant force on a tracked grain as it was entrained from an upstream region until the dune crest, and it is representative of the resultant force acting on grains migrating toward the crest. In Fig. \ref{fig:tracking_force}, the solid blue line corresponds to the longitudinal component of the force, $F_{px}$, and the dashed red line to the transverse component of the force, $F_{pz}$. We observe that the grain experiences a varying force, in the transverse direction the instantaneous values varying around a zero average, while in the longitudinal direction the force has a positive average, with high peaks occurring in some occasions. These peaks are much higher than the mean value (in Fig. \ref{fig:tracking_force}, the highest peak is one order of magnitude greater than the mean value). It is interesting to note that this specific grain took a time slightly higher than 1.5$t_c$ to reach the dune crest, with the highest peak occurring just before reaching the crest.

\begin{figure}[h!]
	\begin{minipage}[c]{\columnwidth}
		\begin{center}
			\includegraphics[width=.85\linewidth]{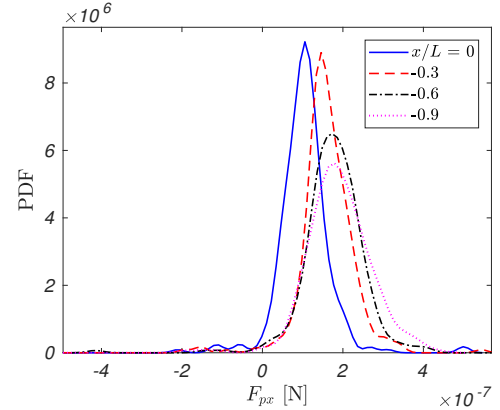}\\
			(a)
		\end{center}
	\end{minipage} \hfill
	\begin{minipage}[c]{\columnwidth}
		\begin{center}
			\includegraphics[width=.85\linewidth]{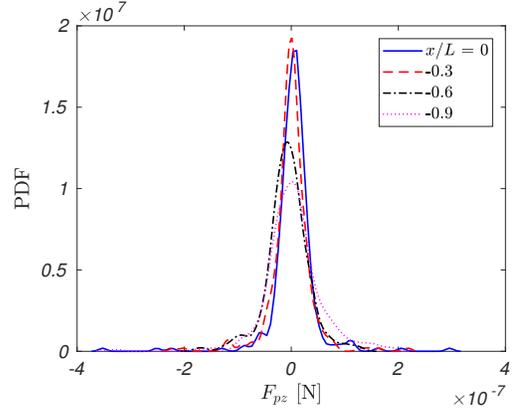}\\	
			(b)
		\end{center}
	\end{minipage}
\caption{PDFs of (a) the longitudinal and (b) transverse components of the resultant force acting on each grain crossing a given barchan cross section. The longitudinal positions of each considered cross section ($x$ planes) are listed in the figure key and $Re$ = 1.47$\times$10$^4$.}
	\label{fig:resultant_forces}
\end{figure}

Figures \ref{fig:resultant_forces}(a) and \ref{fig:resultant_forces}(b) present, respectively, PDFs of the longitudinal and transverse components of the resultant force acting on each grain crossing a given barchan cross section. Figure \ref{fig:resultant_forces} shows that the resultant force acting on each grain has a longitudinal component that attains higher values on upstream regions and decays toward the crest, with a slightly broader distribution upstream, while the transverse component is always peaked at zero, also with a broader distribution upstream. Although the longitudinal component of the resultant force is lower at the crest, the granular flux increases toward the crest, as shown in Subsection \ref{sec:mass_flux}, due probably to the grain inertia. We expect that at the crest the granular flux is lower than that just upstream of it, but, as explained in Subsection \ref{sec:mass_flux}, the relatively high dispersion in computed fluxes does not allow us to prove that for the moment.

The results on the resultant force, which have until now been only conjectured, help to explain the mechanism of upstream erosion and crest deposition that exists on barchan dunes.

\section{\label{sec:Conclu} CONCLUSIONS}

This paper investigated numerically the formation and evolution of single barchans by using computational fluid dynamics - discrete element method (CFD-DEM), where DEM was coupled with large eddy simulation (LES). Our simulations reproduced numerically the experiments of Alvarez and Franklin \cite{Alvarez, Alvarez3}, performed in a closed-conduit channel where initially conical heaps evolved to single barchans under the action of a turbulent water flow. The numerical setup used the same initial and boundary conditions of the experiments of Refs. \cite{Alvarez, Alvarez3}.

Our simulations captured well the evolution of the initial pile toward a barchan dune in both the bedform and grain scales. Concerning the bedform, the morphology obtained from our numerical simulations showed the same characteristic time and length observed in experiments: the horns grow and the dune width and length evolve until $t$ = $2.5 t_c$, where $t_c$ is a timescale proposed in Ref. \cite{Alvarez}, reaching a stable value for $t$ $>$ $2.5 t_c$. The evolution of the horn length, its final value, and that of the dune width and length are in excellent agreement with the experiments of Refs. \cite{Alvarez, Alvarez3}.

The water flow over a fixed barchan was compared with similar experiments reported in Charru and Franklin \cite{Franklin_11}. The evolution of second order moments along the dune obtained from our LES simulations are in good agreement with the results of Ref. \cite{Franklin_11}, showing that the numerical setup adopted is able to capture the mean features associated with the turbulent flow over the dune. 

We tracked the grains that migrated to the growing horns of barchans. We observed a behavior similar to that observed in Ref. \cite{Alvarez3}, with grains that migrate to horns describing circular paths. Plots in polar coordinates centered at the dune centroid showed that these grains come from upstream regions on the periphery of the initial pile. In addition, computations of the total transverse and longitudinal distances traveled by the grains showed PDFs similar to experimental values and a ratio between the transverse and longitudinal distances of approximately 0.5, the same value obtained experimentally.

Finally, we obtained the local granular flux and the resultant force acting on each grain, the latter not yet previously measured nor computed. Concerning the granular flux, we showed that there exists a transient for $t/t_c$ $\lesssim$ 1.5 and that it is not saturated along the barchan. For the resultant force on each grain, we showed that its longitudinal component attains higher values on upstream regions and decays toward the crest. These results help to explain the mechanism of upstream erosion and crest deposition that exists on barchan dunes.

The good agreement between the numerical and experimental results shows that the present method is appropriate to numerical computations of bedforms, opening new possibilities for accessing data not available from current experiments, such as, for example, the instantaneous forces on each grain within the bed.

\section{\label{sec:Ack} ACKNOWLEDGMENTS}

Carlos A. Alvarez is grateful to SENESCYT (Grant No. 2013-AR2Q2850) and to CNPq (Grant No. 140773/2016-9). Erick M. Franklin is grateful to FAPESP (Grant No. 2018/14981-7), to CNPq (Grant No. 400284/2016-2) and to FAEPEX/UNICAMP (Grant No. 2112/19) for the financial support provided. Part of the work was completed while C. A. Alvarez was a visiting graduate student at Western University, Canada, supported through the Emerging Leaders in the Americas Program, Global Affairs Canada. The authors would like to thank Prof. J. M. Floryan from Western University, Canada, for helpful discussions. SHARCNET (www.sharcnet.ca) provided part of computational resources used in the project.

\bibliography{references}

\end{document}